# Significant improvement of flux pinning and irreversibility field in *nano*-Carbon doped MgB$_2$ superconductor

Monika Mudgel[a], V. P. S. Awana[a,*], H. Kishan[a], and G. L. Bhalla[b]

[a]Superconductivity and Cryogenics Division, National Physical Laboratory, Dr. K.S. Krishnan Marg, New Delhi-110012, India
[b]Deptartment of Physics and Astrophysics, University of Delhi, New Delhi-110007, India

Abstract

We report the synthesis and variation of superconductivity parameters such as transition temperature $T_c$, upper critical field $H_c$, critical current density $J_c$, irreversibility field $H_{irr}$ and flux pinning parameter ($F_p$) for the MgB$_{2-x}$C$_x$ system with *nano*-Carbon doping up to x=0.20. Carbon substitutes successfully on boron site and results in significant enhancement of $H_{irr}$ and $J_c(H)$. Resistivity measurements reveal a continuous decrease in $T_c$ under zero applied field, while the same improves remarkably at higher fields with an increase in *nano*-C content for MgB$_{2-x}$C$_x$ system. The irreversibility field value ($H_{irr}$) is 7.6 & 6.6 Tesla at 5 and 10K respectively for the pristine sample, which is enhanced to 13.4 and 11.0 Tesla for x = .08 sample at same temperatures. Compared to undoped sample, critical current density ($J_c$) for the x=0.08 *nano*-Carbon doped sample is increased by a factor of 24 at 10K at 6 Tesla field.

Keywords: A. MgB$_2$ Superconductor, D. Critical current density, D. Flux pinning

PACS No. 74.70.-b, 74.72.-h

*: Corresponding Author
Dr. V.P.S. Awana
Room 109,
National Physical Laboratory, Dr. K.S. Krishnan marg, New Delhi-110012, India
Fax No. 0091-11-25626938: Phone no. 0091-11-25748709
e-mail-awana@mail.nplindia.ernet.in: www.freewebs.com/vpsawana/

Introduction

Due to high critical transition temperature $T_c$ and large coherence length (around 5-10 *nm*), $MgB_2$ possess a great potential for wide engineering and technological applications in comparison to conventional superconductors like $Nb_3Sn$ and Nb-Ti alloy [1] and HTSC compounds [2]. The critical current density of pure $MgB_2$ falls rapidly in high fields because of poor grain connectivity and a lack of pinning centers. But fortunately the flux pinning forces can be enhanced significantly through the doping of *nano*-particles in bulk $MgB_2$ [3-5]. Enormous efforts have been directed in last five years, towards the improvement of superconducting performance of $MgB_2$ through substitution/addition of several *nano*-elements, *nano*-oxides, carbohydrates and other organic compounds [3-5]. The most important role is played by Carbon and its derivatives. The impact of Carbon doping on the superconducting properties especially crystal chemistry and miscibility of Carbon in $MgB_2$ lattice has already been studied in earlier years [6-11]. The solubility of Carbon varies from 2.5% to 30% [7-9]. But from the application point of view, the effect of Carbon doping on flux pinning and critical current density in $MgB_2$ must be appreciated. In the recent years, improvement of $J_c$ is observed through various Carbon sources like diamond, Carbon *nano*-horns, graphite & Carbon *nano*-tubes [12-14].

In this article we explore the effect of amorphous Carbon *nano*-particle substitution in bulk polycrystalline $MgB_2$. Carbon is the only element, which substitutes at Boron site in $MgB_2$ lattice. The partial substitution of Boron by Carbon creates disorder in the superconducting condensate (Sigma band), which is mainly formed by the Boron networks, and enhances the upper critical field value & $J_c(H)$ via intrinsic pinning. Further, remaining unreacted *nano*-Carbon can act as a *nano* additive in the host matrix, causing extrinsic pinning of flux lines. Both ways (substitution/addition) of Carbon/*nano*-Carbon in $MgB_2$ is supposed to enhance the upper critical field, $J_c(H)$ & $H_{irr}$ (irreversibility field) parameters. Recently, some of us elaborated the role of C and its derivatives in superconducting performance of $MgB_2$ superconductor [15]. In present article, we study the high field (up to 14 Tesla) superconducting performance of *nano*-C doped $MgB_2$. The critical current density is improved by an order of magnitude at elevated fields in bulk *nano*-C doped $MgB_2$ samples. Instead of widely used C in $MgB_2$- $_x$C$_x$, we used *nano*-C, which could trigger both intrinsic and extrinsic pinning in the system and as a result spectacular increase in various superconducting parameters. Current article deals mainly with the high field performance of the bulk *nano*-C doped MgB$_2$ samples.

Experimental

The polycrystalline MgB$_{2-x}$C$_x$ samples were prepared by solid-state reaction method at ambient pressure. The starting materials Mg powder (99% pure, 325 meshes), amorphous Boron powder (95% pure, sub micron size) and *nano*-Carbon (10-20*nm*) taken in stoichiometric ratios were mixed thoroughly by continuous grinding. The palletized samples were sintered at a temperature of 850°C for 2.5 hours in Argon atmosphere followed by natural cooling in the same atmosphere. The crystal structure was investigated by *X*- ray diffraction patterns using Rigaku RINT2200HF-Ultima diffractometer with Ni Filtered CuK$_\alpha$ radiation at 40KV and 50mA. Resistivity measurements were made using four-probe technique. Magnetization measurements were carried out on Quantum Design PPMS equipped with VSM attachment.

Results and Discussions:

Figure 1 depicts the X-ray diffraction patterns of series MgB$_{2-x}$C$_x$ (x=0.0 to 0.20) and of *nano*-C in the angular range 20° ≤ 2θ ≤ 70°. The characteristic peak for *nano*-Carbon shown in the top layer does not appear in the Carbon doped samples. All samples of MgB$_{2-x}$C$_x$ series crystallize in hexagonal structure with space group *P6/mmm* [7-11]. All characteristic peaks for the pristine sample are recorded at their well-defined positions and are indexed in the figure. A minor amount of MgO is also determined by the presence of a very low intensity peak at 2θ=63.5° [16]. MgO peak is marked with symbol # in the figure. The peak (100) shifts slightly towards higher angle side resulting in a decrease in *a* parameter. For pristine sample, *a*=3.085Å, while it is decreased to 3.067Å for x=0.20 Carbon doped sample. The (002) peak responsible for *c* parameter is not shifted appreciably and hence *c* parameter is nearly unchanged. Change in lattice parameters confirms the substitution of Boron by Carbon in the lattice. No multiphase formation is noticed up to x=0.10 *nano*-Carbon doped samples. Beyond x=0.10 some extra peaks of MgB$_2$C$_2$ (marked by * in figure 1) appear as impurity phases.

The superconducting transition zone of R(T)-H curves for pure $MgB_2$, x=0.08 and x=0.10 samples are shown in figure 2. The transition temperature ($T_c$) of pure sample is 38.15K, while the same is 34.68K & 34.35K for x=0.08 & x=0.10 samples at zero field. With increment in applied field, resistance curves shift towards lower temperature for both doped & undoped samples. Interestingly the relative shift in $T_c$ with H is much lesser in case of doped sample than the pure one. For example $T_c$ of pure $MgB_2$ sample is 18.75K under 8T applied field, while the same is increased to 20.71 and 20.1K for x=0.08 & x=0.10 samples respectively. Hence substitution of *nano*-Carbon improves the superconducting performance of bulk $MgB_2$ samples at elevated fields. The improvement in critical field values will be discussed later.

The extended magnetization hysteresis loops are shown for doped and undoped samples in both increasing and decreasing field directions at 5, 10 and 20 K in figure 3. The *M-H* loop for doped samples closes much later than the pure sample, which clearly demonstrates the enhanced value of irreversibility field ($H_{irr}$). At 10K, the loop closes at about 6.6 Tesla for the pure sample but is still open upto 11.0 Tesla for the x=0.08 sample. To have a clear picture for all studied samples, the $H_{irr}$ (irreversibility field) are estimated for all samples at 5, 10 and 20K from their respective magnetization loops. $H_{irr}$ is taken as the applied field value at which magnetization loop almost closes with a criterion of giving critical current density value of the order of $10^2$ A/cm$^2$. The variation of irreversibility field values with doping content is depicted in figure 4 at different temperatures. The x=0.08 sample has highest $H_{irr}$ values among all the samples. For pristine sample, the $H_{irr}$ values are 4.3, 6.6 & 7.6 Tesla at 20,10 & 5K respectively, whereas it is increased to 6.3, 11.0 & 13.4 Tesla for the x=0.08 sample at the same temperatures. These values are either slightly higher or in confirmation with those as reported earlier by Soltanian et al [17]. The increased values of $H_{irr}$ confirm the flux pinning by added *nano*-Carbon particles.

Critical field values vs reduced transition temperature curves are plotted in the inset of figure 4 for pure and doped samples. The magnetic field value required to destroy the surface superconductivity is more than the field required to destroy superconductivity in bulk [18, 19]. In fact the former is about 1.7 times the latter one near transition temperature. Although the ratio of two denoted as η varies for a two-band superconductor with temperature, but it attains its maximum value near transition temperature [19]. So, critical field values are calculated from the $R(T)$-$H$ curves with the criteria; $H_c = 1.7 \times H$, where $H$ is the applied field value at which $R \rightarrow 0$. All the doped samples have higher upper critical field values ($H_c$) than pure sample at all temperatures. The x=0.08 & x=0.10 sample have almost similar values of upper critical field up to 6Tesla and the former is better than the later for higher applied fields. The separation of un-doped sample curve from the doped ones demonstrates the inferior superconducting performance of pure $MgB_2$ than the *nano*-Carbon doped samples at higher fields. The Carbon atom is substituted at the Boron site in host $MgB_2$, which has one more electron than boron. This extra electron is donated to the sigma band creating disorder in it, which is responsible for enhanced upper critical field values.

The critical current density is calculated from the magnetization hysteresis loops using Bean's Critical Model. The variation of $J_c$ with applied fields is shown in figure 5 for doped & undoped samples at 10K. All samples have $J_c$ of the order of $10^5$ A/cm$^2$ at low fields. For higher fields, quantitatively, $J_c$ is about $1.13 \times 10^4$ A/cm$^2$ at 6Tesla and 10K for x=0.08 *nano*-Carbon doped sample, where as it is $4.58 \times 10^2$ A/cm$^2$ for pure sample at same field and temperature values. More specifically, $J_c$ of this sample is 24 times higher than the pure sample at 6 Tesla & 10K. The similar is the situation at 20K as well for the $J_c(H)$ of x=0.08 sample, which is 14 times higher than that of the pure sample at a field value of 4.0 Tesla, which is shown in inset of figure 5. The critical current density value is enhanced much profoundly in the case of x=0.08 sample at both the temperatures, proving it to be an optimal composition for this batch of samples. The observed values of $H_c$, $H_{irr}$ and $J_c$ are competitive or slightly better than those being reported yet [20-24].

In order to have a confirmative claim about the flux pinning behavior, flux pinning force $F_p$ is calculated. The dependence of reduced flux pinning force with applied field at 10K is demonstrated in figure 6. The relationship between flux pinning force and critical current density could be described by [25]

$$F_p = \mu_0 J_c(H) H \quad (1)$$

where $\mu_0$ is the magnetic permeability in vacuum. The curves become broader with *nano*-Carbon doping as compared to the undoped sample. It is due to pinning, which results in increased critical field values and subsequently enhanced values of critical current density. The shape of flux pinning curve is explainable by grain boundary/point defect pinning. But a shift in the peak is also noticed (by about 1 Tesla) for the doped samples in comparison to the pure one (marked with arrows in Fig. 6). Reportedly there are three main pinning mechanisms viz., grain boundary pinning, point defect pinning (seen as broadening of pinning plot) and order parameter change (seen as peak shift in pinning plot) [26]. In our case the shift in the pinning plot peak by over one Tesla clearly demonstrates the disturbance to order parameter. Though most of the nominally substituted *nano*-Carbon has gone to B-site, the remaining minor quantity (marked as * in Fig.1) could yet introduce point defects/ grain boundary pinning in the $MgB_2$ matrix. Both intrinsic (B-site *nano*-C substitution) and extrinsic *nano*-Carbon derivatives (marked * in Fig.1) pinning are responsible for spectacular superconducting performance of the studied $MgB_{2-x}C_x$ system.

Although x=0.08 sample seems to show best results for this batch of samples, but we can not regard it as an optimum composition, because depending on some other factors like precursor material and processing conditions, the optimum Carbon content for best performance may change slightly. That's why different groups reported different compositions (x=0.04 to 0.15) showing best performance [17-22]. Even in our different batch of samples [15], the x=0.10 sample showed best results up to a field of 8 Tesla. The major fact to be highlighted is that *nano*-Carbon addition results in enhancement of $H_c$, $H_{irr}$ and $J_c$ values. All doped samples show better performance than undoped sample.

Conclusion

The effect of *nano*-Carbon doping is studied in relation to the enhancement of superconducting performance of widely known $MgB_2$ superconductor. The main thrust is given on the magnetization and flux pinning properties. The $H_c$, $J_c$ & $H_{irr}$ values obtained are competitive or slightly better than the reported values on doped bulk $MgB_2$ superconductor. The sample with x=0.08 in $MgB_{2-x}C_x$ series proved to be best one, which possess a quiet high value of $H_{irr}$=13.4 Tesla at 5K. Along with the improvement in $H_{irr}$ value, $H_c$ value is also enhanced much more profoundly. $J_c$ value is increased by more than an order of magnitude for x=0.08 sample. Our values for *nano*-Carbon doped $MgB_2$ are better than as reported for normal C doping in this system.

Acknowledgement

The authors from *NPL* would like to thank Dr. Vikram Kumar (*DNPL*) for his great interest in present work. Dr. Rajeev Rawat from *CSR*-Indore is acknowledged for the *R (T) H* measurements. Mr. Kranti Kumar and Dr. A. Banerjee are acknowledged for the high field magnetization measurements. Monika Mudgel would like to thank the *CSIR* for JRF award to pursue her *Ph. D* degree.

Figure Captions

Figure 1. X-ray diffraction patterns for the $MgB_{2-x}C_x$ series (x=0.0, 0.04, 0.08, 0.10, 0.15 & 0.20) and *nano*-Carbon. The symbol # marks the minor presence of MgO

Figure 2. Superconducting transition zone of normalized Resistance vs Temperature plot *R (T) H* up to 8 Tesla for pure, x=0.08 & x=0.10 Carbon doped samples

Figure 3. Magnetization loop *M (H)* at 10K for $MgB_{2-x}C_x$ samples (x=0.0, 0.04, 0.08, 0.10 & 0.20) up to 12 Tesla field

Figure 4. Irreversibility field $H_{irr}$ versus Carbon content plots at 5, 10 & 20K for $MgB_{2-x}C_x$ samples. The inset shows the upper critical field ($H_c$) vs Normalized temperature plots for $MgB_{2-x}C_x$ samples (x=0.0, 0.08, 0.10 & 0.20)

Figure 5. $J_c(H)$ plots for $MgB_{2-x}C_x$ samples (x = 0.0, 0.04, 0.08, 0.10 & 0.20) at 10K. The inset shows the same at 20K

Figure 6. Variation of reduced flux pinning force ($F_p/F_{p,max}$) with magnetic field for $MgB_{2-x}C_x$ samples (x=0.0, 0.04, 0.08, 0.10 & 0.20) at 10K. The inset shows the same at 20K. The arrows mark the shift in pining peaks.

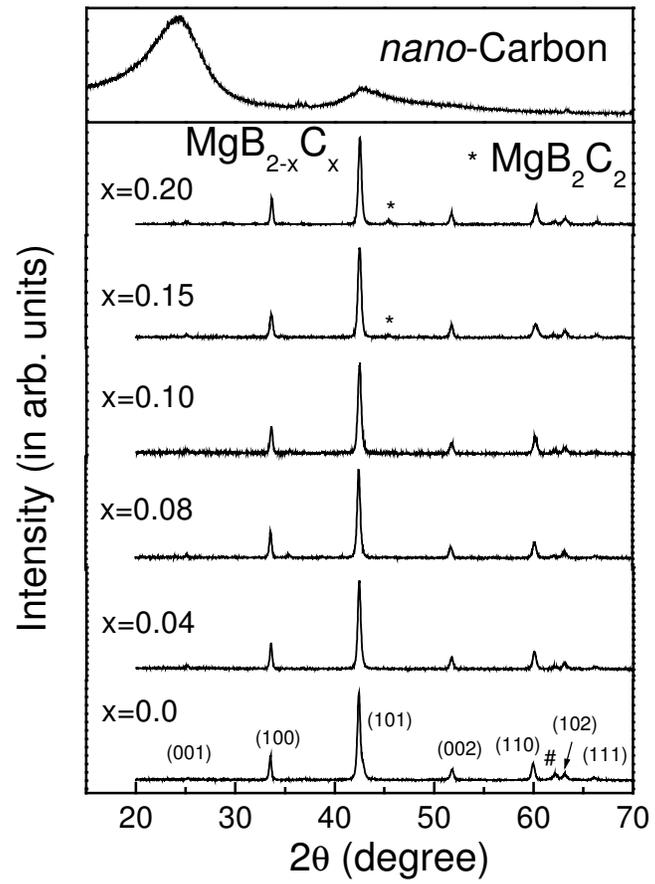

Fig. 2

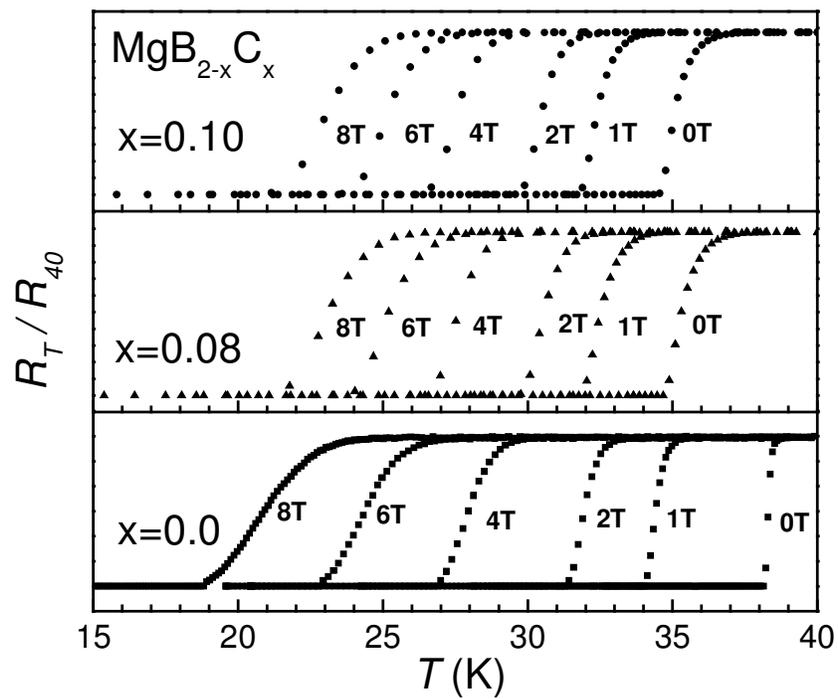

Fig. 3

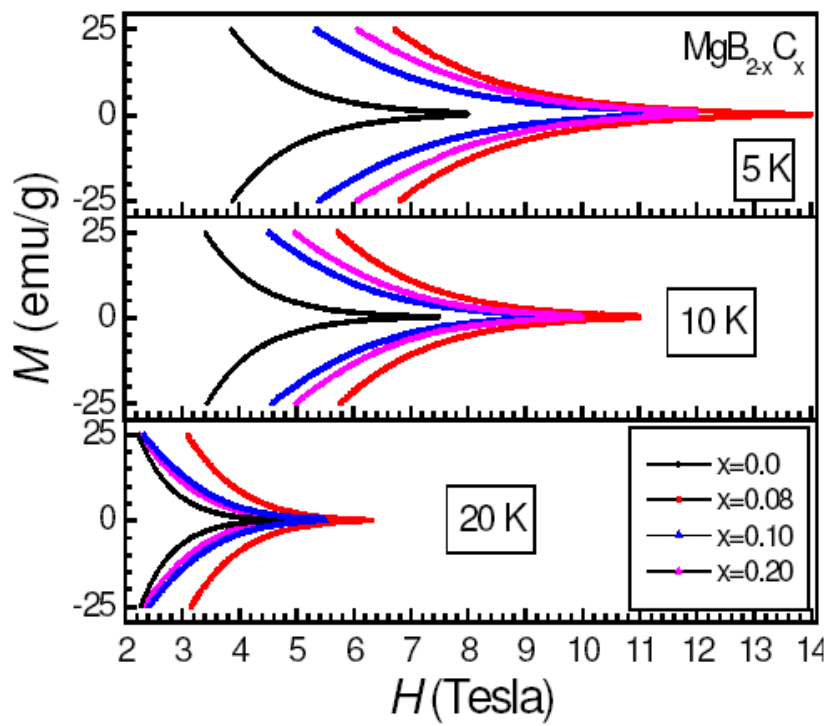

Fig. 4

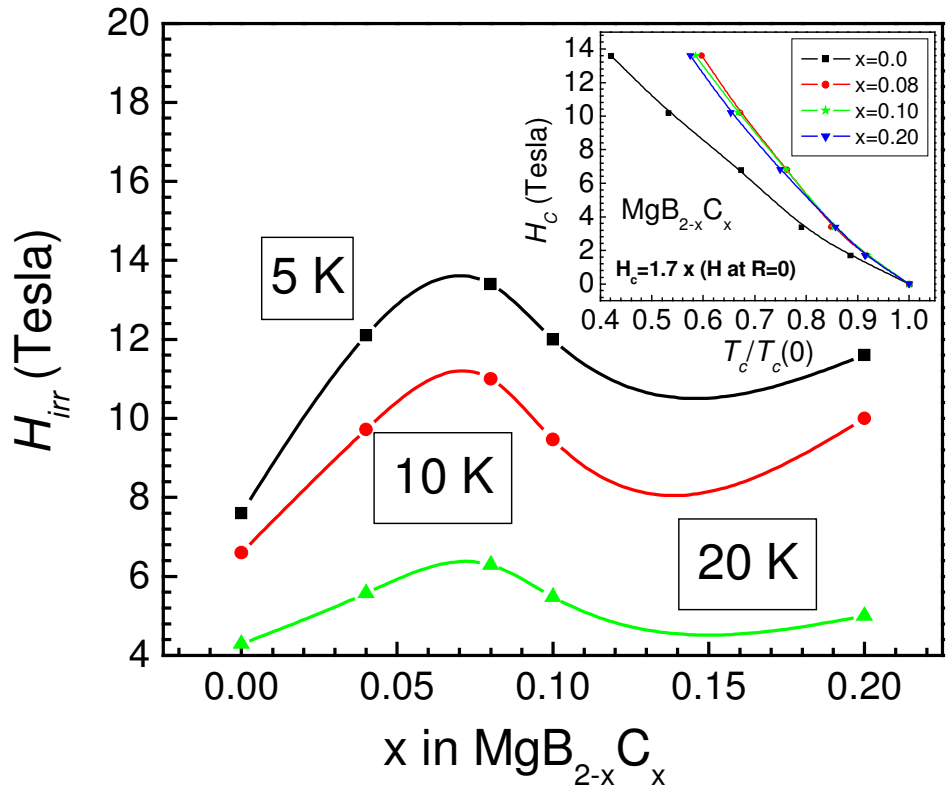

Fig. 5

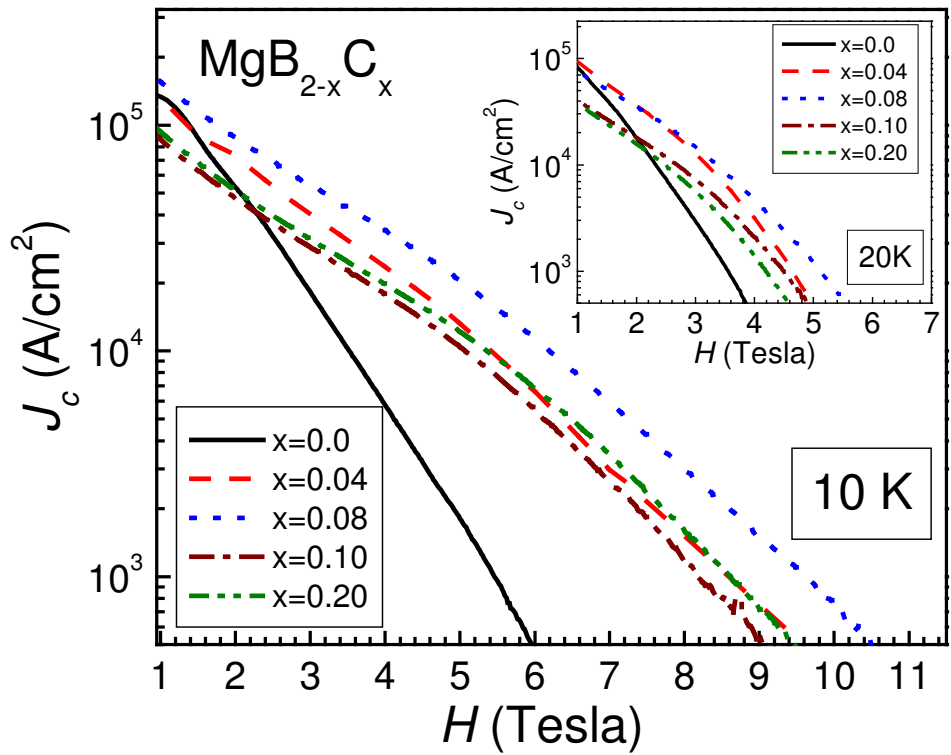

Fig.6

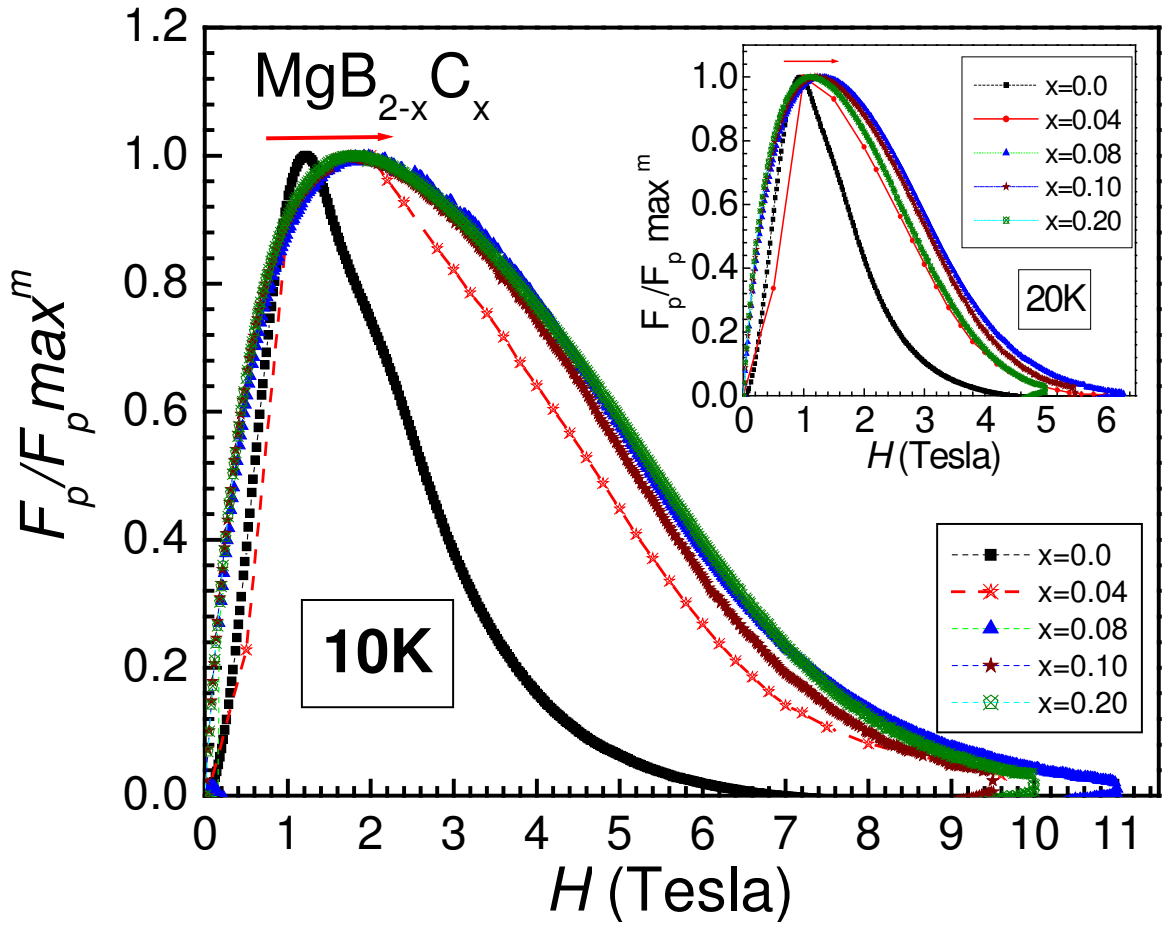